\documentclass[9pt,twocolumn,twoside]{osajnl}

\journal{ol} 

\setboolean{shortarticle}{true}

\title{Polarization diversity phase modulator for measuring frequency-bin entanglement of a biphoton frequency comb in a depolarized channel}

\author[1,$\dagger$]{Oscar E. Sandoval}
\author[1,$\dagger$,*]{Navin B. Lingaraju}
\author[1]{Poolad Imany}
\author[1]{Daniel E. Leaird}
\author[2]{Michael Brodsky}
\author[1]{Andrew M. Weiner}

\affil[$\dagger$]{These authors contributed equally to this work.}
\affil[1]{School of Electrical and Computer Engineering, Purdue University, 465 Northwestern Ave, West Lafayette IN 47901, USA}
\affil[2]{U.S. Army Research Laboratory, Adelphi, MD 20783, USA}

\affil[*]{nlingara@purdue.edu}

\begin{abstract}
Phase modulation has emerged as a technique to create and manipulate high-dimensional frequency-bin entanglement. A necessary step to extending this technique to depolarized channels, such as those in a quantum networking environment, is the ability to perform phase modulation independent of photon polarization. This also necessary to harness hypertanglement in the polarization and frequency degrees of freedom for operations like Bell state discrimination. However, practical phase modulators are generally sensitive to the polarization of light and this makes them unsuited to such applications. We overcome this limitation by implementing a polarization diversity scheme to measure frequency-bin entanglement in arbitrarily polarized photon pairs.
\end{abstract}

\setboolean{displaycopyright}{true}

\begin{document}

\maketitle

Polarization-entangled photons are a popular choice for quantum networking protocols owing to their compatibility with standard telecommunications equipment. Full Bell state discrimination is a prerequisite for superdense coding \cite{Barreiro2008} and teleportation \cite{graham2015}. However, entanglement in one degree of freedom alone is insufficient to perform a full Bell state analysis using linear optics. Entanglement in a second degree of freedom has been used to resolve this limitation \cite{Walborn2003,Barreiro2008}. Other degrees of freedom, like orbital angular momentum and time-bin, for example, have been used to carry out full Bell state discrimination \cite{Barreiro2008,Schuck2006}.  

However, entanglement in the frequency degree of freedom, which is compatible with modern fiber optic networks, has largely been untapped as a resource for quantum communication. Recently there has been work outlining ways to perform quantum information processing in the spectral domain using only phase modulators and Fourier transform pulse shapers \cite{lukens2017}. Furthermore, on-chip microresonators have been shown to be an excellent source for generating photon pairs through spontaneous four wave mixing \cite{Kues2017,imany2018}. One state that is of particular interest is the biphoton frequency comb (BFC) - a coherent superposition state consisting of N-energy matched comb line pairs. If $\alpha_{k}$ represents the complex amplitude of the $k^{th}$ comb line pair, the general state of a BFC can be written as \cite{Imany2018pra}: 

\begin{equation}
\left | \Psi  \right >_{BFC} = \sum_{k=1}^{N} \alpha_{k} \left | k,k \right > _{SI}
\label{bfc_state}
\end{equation}

Recent work has shown that electro-optic phase modulation can be used to mix frequencies from different modes and carry out two-photon interferometry that is sensitive to the phase on comb line pairs \cite{Kues2017,imany2018,Imany2018pra}, thereby establishing high-dimensional frequency-bin entanglement. However, phase modulators - whether based on the linear electro-optic effect in $\chi^{(2)}$ materials, or on the carrier dispersion effect in silicon, or on resonant cavities - are sensitive to the polarization of light. In the case of devices based on lithium niobate \cite{wooten2000,wang2018}, only the $r_{33}$ electro-optic coefficient has a magnitude large enough to generate a meaningful change in the waveguide effective index. In the case of silicon modulators \cite{fukuda2008} or resonant-cavity plasmonic modulators \cite{haffner2018}, large waveguide birefringence results in polarization mode dispersion and polarization-dependent loss. Therefore, harnessing hyperentanglement in the polarization and frequency degrees of freedom, for use over fiber-optic networks, requires a polarization diversity scheme capable of measuring frequency-bin entanglement irrespective of photon polarization. 

\begin{figure}[ht!]
\centering\includegraphics[width=0.65\linewidth]{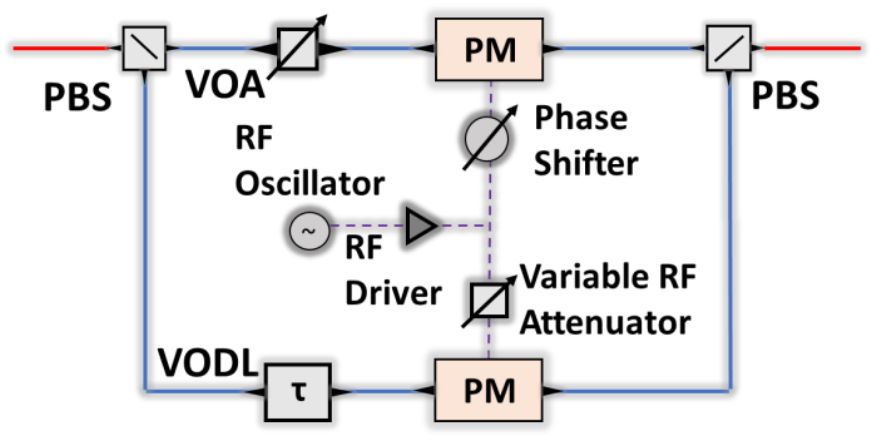}
\caption{Schematic of polarization diversity phase modulator (single-mode and polarization maintaining fiber are depicted in red and blue, respectively).}
\label{pdpm_setup}
\end{figure}

The scheme used to achieve polarization-independent phase modulation is shown in figure \ref{pdpm_setup}. The polarization diversity phase modulator (PDPM) comprises two fiber-based polarization beam splitters (PBS) and two phase modulators (PM). Light enters the PDPM through the PBS on the left and is split into orthogonally polarized components that propagate along separate channels. Both channels of the device use polarization maintaining fiber and light in each channel is modulated independently. A second fiber PBS recombines light from the two channels to return a phase modulated version of the arbitrarily polarized input signal.

To carry out polarization-independent phase modulation for projective measurements on time-energy entangled photons, the two paths through the PDPM need to be sufficiently indistinguishable that path information cannot be gleaned from the system \cite{Brodsky2011,Jones2018}. Namely, the path length and optical loss through the two arms of the PDPM need to be nearly identical. Furthermore, the RF drive to the two phase modulators - in terms of modulation depth and RF delay - needs to be balanced as well. A variable optical attenuator (VOA) was placed in the arm of the PDPM with lower initial loss and tuned until the optical loss through each arm was the same. To determine the path length difference between the arms of the PDPM, broadband light was launched into the PDPM so that at least some light entered each arm. The output of the PDPM was sampled with a polarizer (oriented at $45^{\circ}$ relative to the polarization in PDPM channels) and sent to an optical spectrum analyzer (OSA). The path length difference between the arms was estimated from the spacing of the spectral fringes \cite{weiner2011}. By suitably adjusting the variable optical delay line (VODL), we reduced the path length difference between the arms of the PDPM. We were able to set the relative delay to zero with a precision of $\approx$60 fs, limited by slow delay fluctuations in the ambient environment. 

Each phase modulator in the PDPM has its own modulation efficiency ($V_{\pi}$). Consequently, when both phase modulators are driven with the same RF waveform and continuous wave optical test signal, frequency combs with slightly different spectra are generated. To ensure that both phase modulators impart identical phase shifts, an RF attenuator was used to adjust the power delivered to the phase modulator with higher modulation efficiency. To minimize the RF delay between the driving waveforms, the output of the PDPM was sampled in a manner similar to that described above. Any delay between the RF driving waveforms induces a linear spectral phase shift, which manifests as an asymmetry in the comb spectrum. An RF phase shifter was used to reduce the RF delay between the driving waveforms. 

As currently constructed, the PDPM has an insertion loss of 5.6dB. However, one of the larger sources of loss stems from the fact that the phase modulators in the PDPM have different insertion losses (2.7dB and 3.7dB). Matching their losses to the lower value would shave a dB off the overall system loss.

\begin{figure}[ht!]
\centering\includegraphics[width=\linewidth]{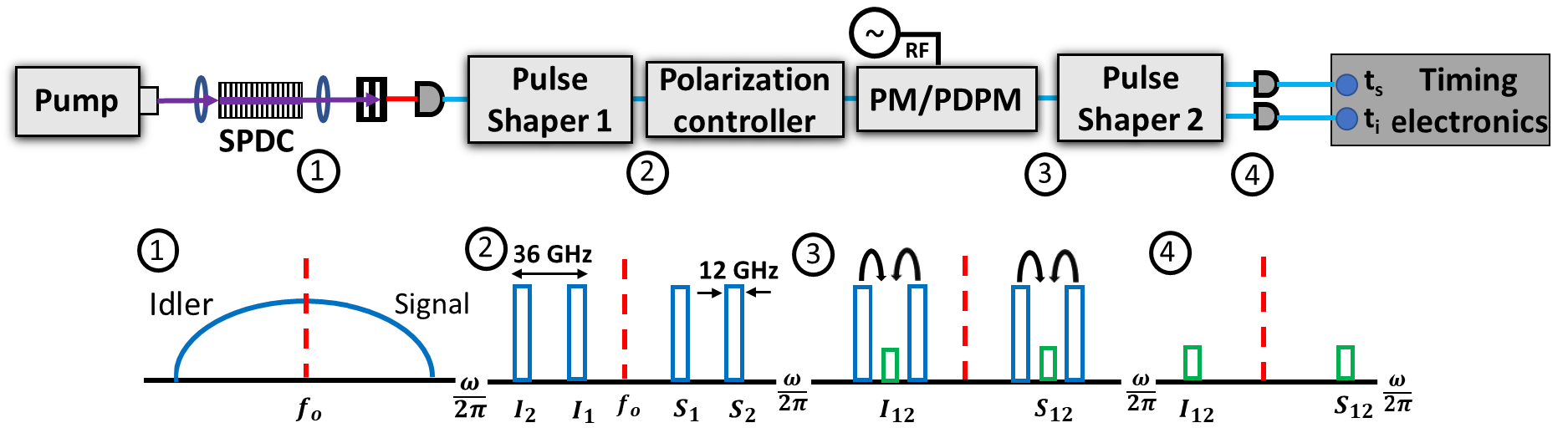}
\caption{Experimental setup for measuring frequency-bin entanglement in a biphoton frequency comb. Note that either a standalone phase modulator or the PDPM is used to mix frequencies in this arrangement.}
\label{fbe_setup}
\end{figure}

After matching both arms of the PDPM, the polarization diversity scheme was used to characterize frequency-bin entanglement in a biphoton frequency comb (BFC) in a manner similar to that presented in references \cite{Kues2017,imany2018,Imany2018pra}. In order to characterize frequency-bin entanglement across different comb line pairs, different frequency modes are mixed using electro-optic phase modulation. If, for example, the frequency of phase modulation equals one half the free spectral range (FSR) of the BFC, one can project adjacent signal and idler pairs on top of each other, thereby creating an indistinguishable superposition state at frequencies halfway between the original modes ($I_{12}$ and $S_{12}$ in figure \ref{fbe_setup}). By varying the joint phase on one of the BFC comb line pairs, it is possible to vary the overall amplitude of this superposition state and, therefore, the probability of detecting coincidences at intermediate frequencies $I_{12}$ and $S_{12}$. The result is a sinusoidal variation in the number of coincidences as a function of the joint phase on one of the comb line pairs.

The setup for carrying out two-photon interferometry and measuring frequency-bin entanglement across comb line pairs is shown in figure \ref{fbe_setup}. A continuous wave laser pumps a periodically poled lithium niobate (PPLN) waveguide, which generates a broadband (5 THz) spontaneous parameteric downconversion (SPDC) spectrum. A BFC is carved from this continuous spectrum by a pulse shaper ("Pulse Shaper 1," Finisar Waveshaper 1000s). Pulse shaper 1 was set to yield a BFC with an FSR of 36 GHz and each frequency mode had a bandwidth of 12 GHz. A polarization controller was placed after the pulse shaper and used to vary the polarization of the BFC at the input of the PDPM (or standalone PM). The PDPM (or standalone PM) was driven at half the FSR of the BFC (18 GHz) in order to overlap adjacent frequency modes. A second pulse shaper ("Pulse shaper 2," Finisar Waveshaper 4000s) was used to demultiplex the output of the PDPM (or standalone PM) and send signal and idler photons at the intermediate frequencies to a pair of InGaAs single photon detectors (Aurea SPD-AT-2). An event timer (HydraHarp 400) was used to tag single photon events and generate a histogram of two-photon coincidences.

\begin{figure}[ht!]
\centering\includegraphics[width=0.7\linewidth] {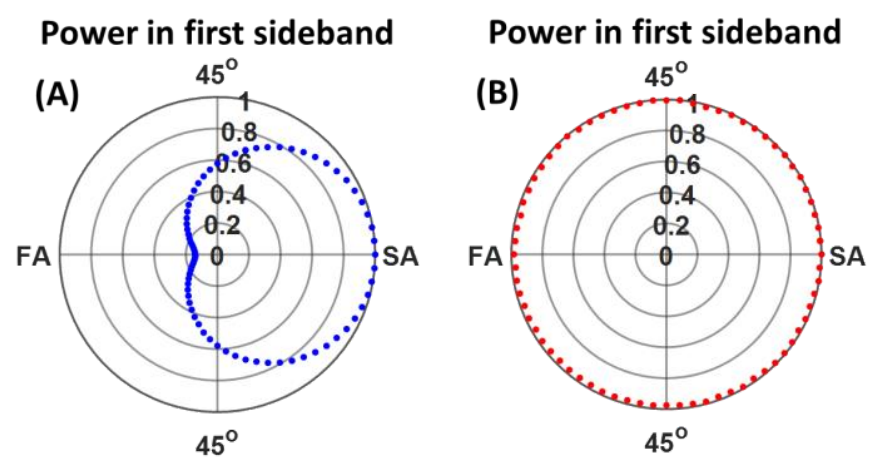}
\caption{Normalized power in the +1 sideband for different orientations of linearly polarized light in the standalone PM (A) and PDPM (B). The polarization state has been defined with respect to the slow (SA) and fast axes (FA) of the modulator fiber input.}
\label{classical}
\end{figure}

To quantify the benefits of using a polarization diversity scheme, the performance of the PDPM was compared with that of a standalone PM. Both devices were first characterized using linearly polarized light from a continuous wave laser. A deterministic polarization controller (DPC) was used to vary the state of linear polarization at the input port of a standalone PM. The results are illustrated in figure \ref{classical}A. The PM efficiently scatters light into the +1 sideband when the state of polarization is aligned with the slow axis (SA in figure \ref{classical}) of the PM fiber input. This is the direction that is aligned with both the $r_{33}$ electro-optic coefficient of lithium niobate and the electric field of the modulating RF waveform. As the polarization is tuned toward the fast axis of the PM fiber input (FA in figure \ref{classical}), some of the light in the PM is now orthogonal to the $r_{33}$ coefficient and passes through the device undergoing substantially reduced phase modulation (power in the +1 sideband is a factor of 0.14 relative to that when the polarization is aligned with $r_{33}$ coefficient). These experiments were repeated with the PDPM and the results are shown in figure \ref{classical}B. The polarization diversity scheme functions as expected and the normalized optical power in the +1 sideband varies over the range from 0.986 to 1 with a standard deviation of 0.4\%.

Two-photon interference measurements were then carried out on a BFC with only two comb-line pairs. In this case, equation \ref{bfc_state} simplifies to: 

\begin{equation}
\left | \Psi  \right >_{BFC} = \frac{1}{\sqrt{2}} \left[ \left|1,1\right>_{SI} + e^{i\left ( \phi_{S2} + \phi_{I2} \right )}\left|2,2\right>_{SI} \right]
\label{bfc_phase}
\end{equation}

\begin{figure}[ht!]
\centering\includegraphics[width=0.7\linewidth]{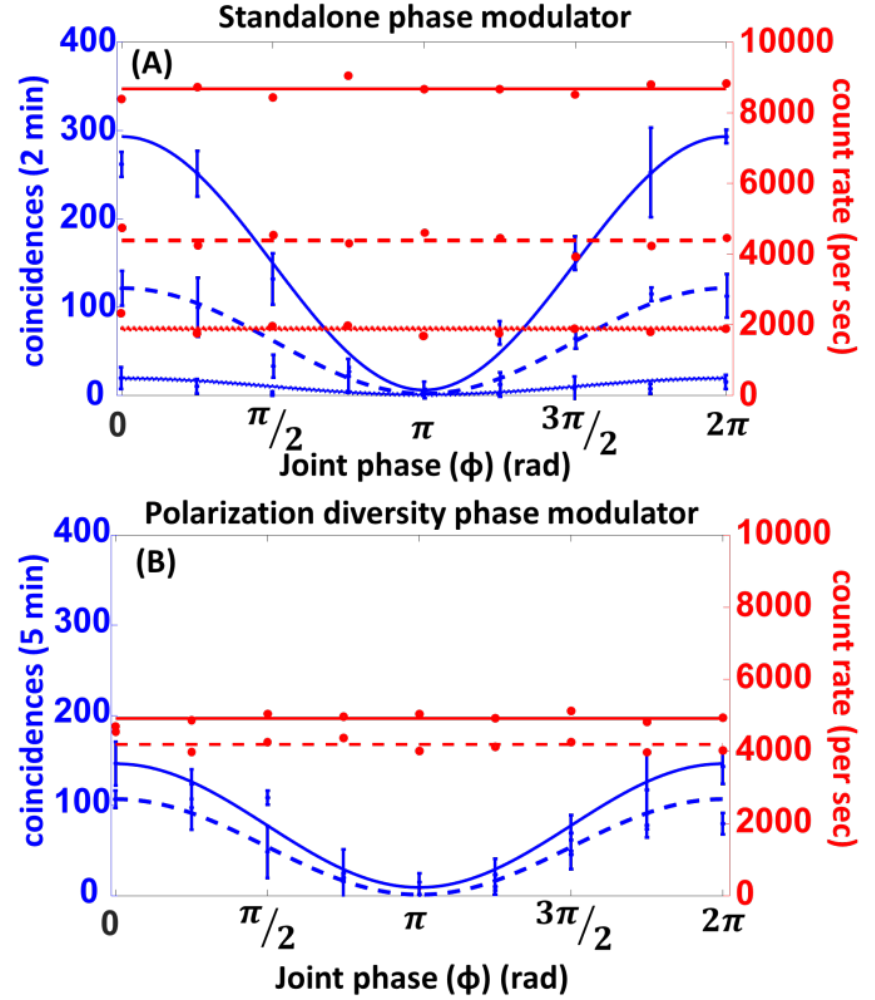}
\caption{Results from frequency-bin entanglement measurements. (A) Coincidences (blue) and counts (red) when using the standalone PM for three different orientations of the polarization - $0^{\circ}$ (solid), $90^{\circ}$ (dotted), and an intermediate orientation (dashed). As the polarization of the BFC is tuned away from $0^{\circ}$, the number of counts at the overlap frequencies ($I_{12}$ and $S_{12}$) falls because the component of the wavefunction aligned with the modulating RF field gets smaller and smaller. (B) Coincidences (blue) and counts (red) when using the PDPM for two cases - one where photons are equally likely to be in either arm (dashed) and the other where photons go entirely through one arm (solid).}
\label{fbe_original}
\end{figure}

For each device - PDPM and standalone PM - interference traces were generated by tracking the number of coincidences, and count rate, as a function of the joint phase ($ \phi_{S_{2}} + \phi_{I_{2}} $) on $\left|2,2 \right>_{SI}$. For each value of the joint phase, coincidences were recorded over three equal time intervals (5 minute intervals for the PDPM and 2 minute intervals for the standalone PM). The average number of coincidences and their standard deviation over these three intervals were calculated after subtracting accidentals. The full interference traces, for 9 different values of the joint phase, are shown in figure \ref{fbe_original}. For the standalone PM, two-photon interference traces were generated for three different polarization states of the BFC - $0^{\circ}$, $90^{\circ}$, and an orientation roughly in between these states. These angles are defined relative to the slow axis of the PM fiber input, which is polarization-maintaining. As the polarization of the BFC is tuned away from $0^{\circ}$, the number of counts at the overlap frequencies ($I_{12}$ and $S_{12}$) is expected to fall as the component of the wavefunction aligned with the modulating RF field gets smaller and smaller. As expected, the number of coincidences falls off at a sharp rate resulting in lower fringe amplitude. When the BFC is aligned with the FA of the PM fiber input (dotted lines in figure \ref{fbe_original}A), the count rate (red) is reduced to 0.28 of that when the BFC is aligned with the SA (solid lines in figure \ref{fbe_original}A). The recorded coincidences (blue) from the BFC are reduced to a factor of 0.07 compared to when the BFC is aligned with the SA. For the PDPM, an interference trace was recorded for the instance when photons were equally likely to end up in either arm of the device (dashed lines in figure \ref{fbe_original}B). To see if there was any residual polarization dependence, an interference trace was then recorded for the instance when all the photons passed through only one arm of the device (solid lines in figure \ref{fbe_original}B). Drift in the setup over the long acquisition time (around three hours) may be responsible for some of the differences between the traces in figure \ref{fbe_original}B. However, the polarization diversity scheme is clearly more robust to changes in BFC polarization. The count and coincidence rates for the PDPM are lower compared to that of the standalone PM (BFC polarization at $0^{\circ}$) because the polarization diversity scheme has a higher insertion loss.

\begin{figure}[ht!]
\centering\includegraphics[width=0.75\linewidth]{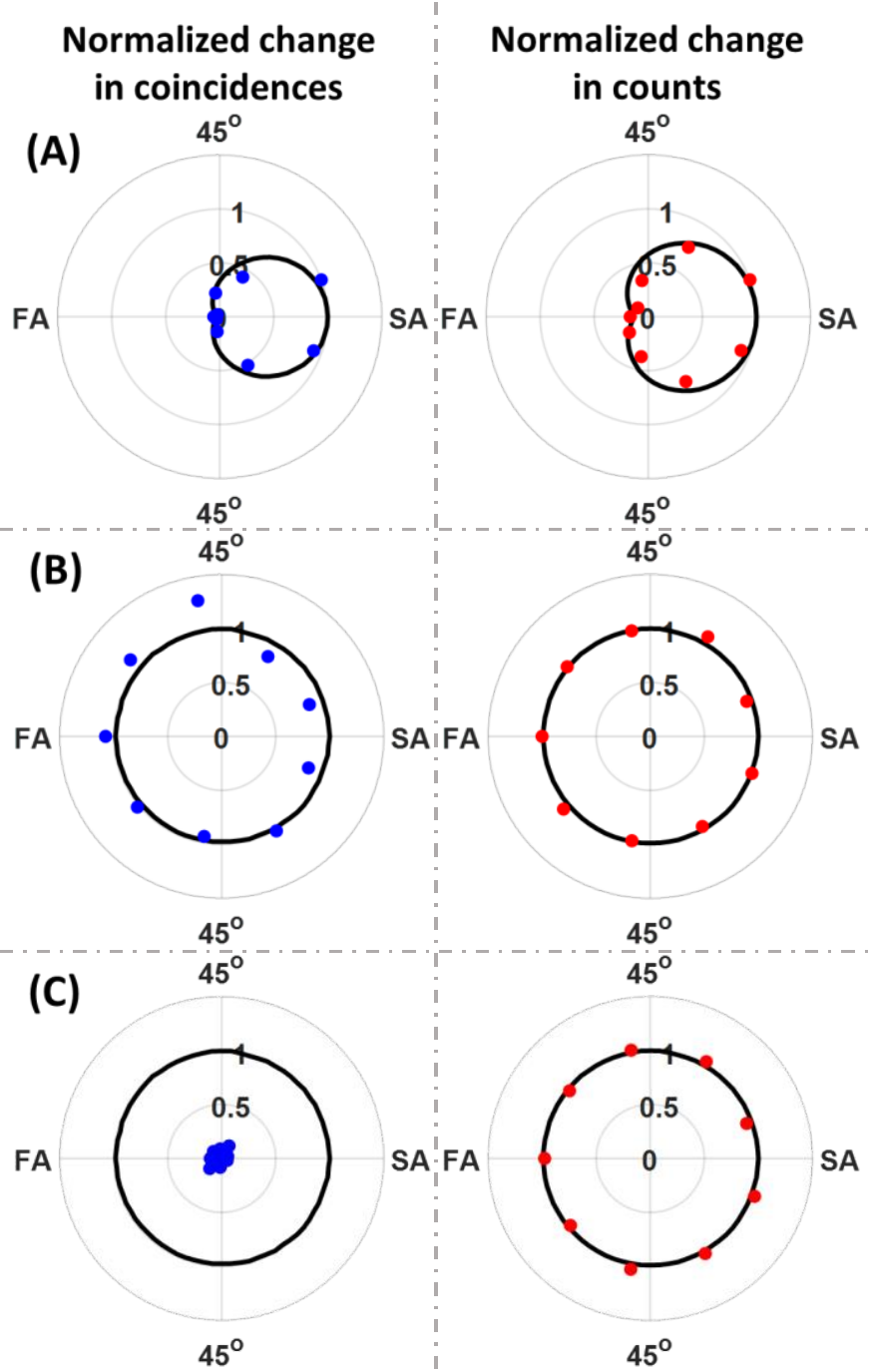}
\caption{(A) Normalized coincidence (blue) and count rates (red), as a function of BFC polarization, from frequency-bin entanglement measurements using a standalone PM. The results are plotted on an azimuthal slide of the Poincar\'e sphere. (B)-(C) are analogous to (A) but for the PDPM. In (A) and (B), the joint phase on $\left|2,2 \right>_{SI}$ was set to $0$. In (C), the joint phase was set to $\pi$. (SA - slow axis and FA - fast axis)} 
\label{fbe_pol}
\end{figure}

Having established that the PDPM can be used to visualize strong two-photon interference in a BFC, irrespective of polarization, we now look to quantify its performance and contrast it with that of a standalone PM. For these measurements, we varied the BFC polarization state using a deterministic polarization controller while keeping the joint phase on $\left|2,2 \right>_{SI}$ set to $0$, which is the value for which coincidences at the overlap frequencies are expected to be at a maximum. The results are presented in figures \ref{fbe_pol}. For measurements with the standalone PM, count and coincidence rates are sensitive to BFC polarization (figure \ref{fbe_pol}A). For measurements with the PDPM, count and coincidence rates are relatively immune to changes in BFC polarization (figure \ref{fbe_pol}B). The results from classical measurements presented in figure \ref{classical} were used to determine the expected change in the count and coincidence rates as a function of BFC polarization. Counts are expected to track with the power scattered into the +1 sideband during classical experiments. Coincidences are expected to fall off quadratically with respect to the drop in power in the +1 sideband. Based on these relationships, the expected change in the normalized count and coincidence rates was plotted (black lines) alongside the actual change in count and coincidence rates in figure \ref{fbe_pol}. There is good agreement between the data from classical measurements and that from two-photon interferometry. Finally, the measurements using the PDPM were repeated, but with the joint phase on $\left|2,2 \right>_{SI}$ set to $\pi$ (figure \ref{fbe_pol}C). Based on the interference traces in figure \ref{fbe_original}, coincidences at the intermediate frequencies are expected to be at a minimum for this value of the joint phase, which is what was observed in this instance as well. These measurements were made in order to establish that the coincidences in figure \ref{fbe_pol}B are the result of two-photon interference and not the result of leakage from the original signal and idler bins. 

Thus far, two-photon interference measurements were performed on co-polarized signal-idler pairs. However, it is important that any polarization diversity scheme be sufficiently robust to also handle instances where the entangled photons are orthogonally polarized with respect to one another. Since a standalone PM can only modulate one polarization component at a time, we expect that its ability to efficiently mix frequencies of orthogonally polarized signal-idler pairs will be limited. 

To prepare orthogonally polarized signal-idler pairs, we took advantage of frequency dependent polarization rotation in a birefringent fiber. Our SPDC source was carved into a BFC consisting of signals and idlers 1.3 THz apart from each other, with the $S_{1}-S_{2}$ and $I_{1}-I_{2}$ spacing kept at 36 GHz. The resulting BFC was launched at $45^{\circ}$ relative to the slow axis of a 28 cm section of polarization-maintaining fiber (with an estimated differential group delay of 1.73 ps/m). The signal-idler spacing was chosen so that the number of Poincar\'e sphere rotations undergone by the signal and idler polarizations differs by one half rotation at the output of the section of polarization-maintaining fiber. Interference traces similar to those recorded in figure \ref{fbe_original} were then generated, but with the signal and idler polarizations orthogonal to one another. As expected, the amplitude of the two-photon trace generated using a standalone phase modulator is very low since a standalone PM is unable to efficiently mix the frequencies of both the signal and the idler photons simultaneously. The PDPM is clearly not limited in this regard.

\begin{figure}[ht!]
\centering\includegraphics[width=0.7\linewidth]{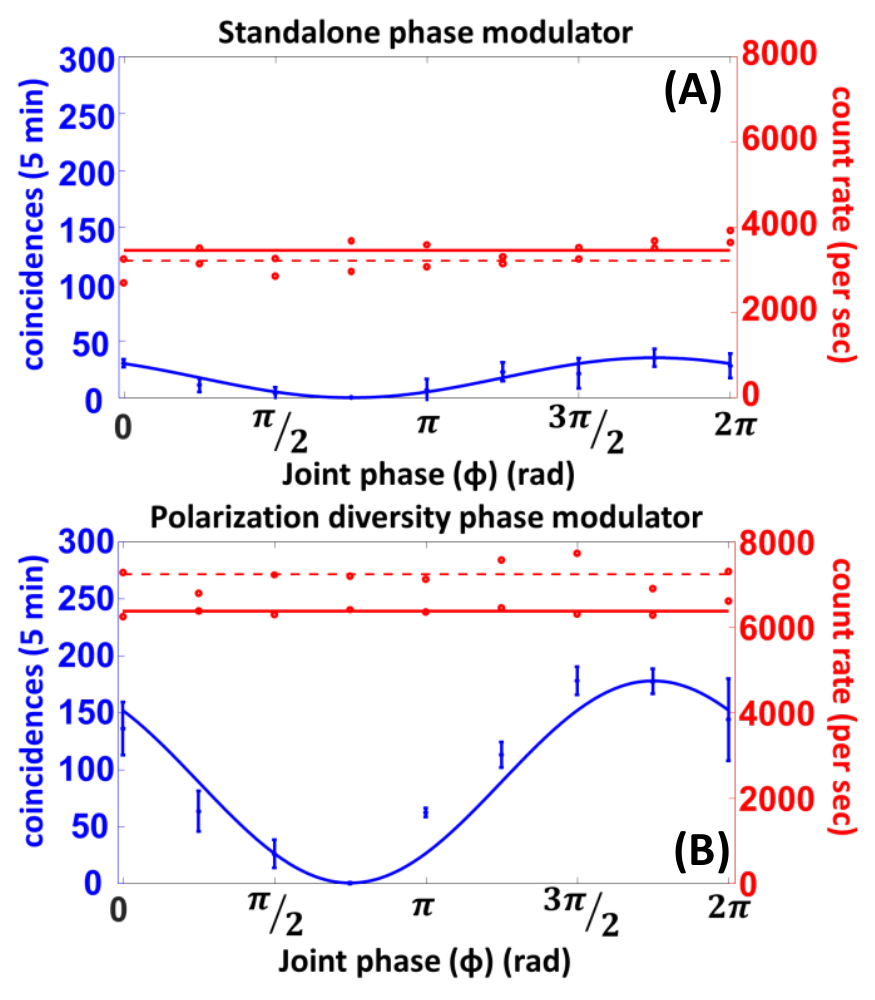}
\caption{Frequency-bin entanglement measurements using orthogonally polarized signal and idler pairs for (A) the standalone PM and (B) the PDPM. Dashed and solid red lines correspond to the signal and idler count rates, respectively.}
\label{fbe_ortho}
\end{figure}

The data presented in this report shows the PDPM to be a promising tool for future quantum networks based on fiber optic infrastructure. However, one limitation of the current polarization diversity scheme is that it does not include a mechanism to stabilize the phase difference between the arms of the PDPM. Consequently, the relationship between the polarization of incoming photons and photons leaving the PDPM fluctuates with time. For applications to quantum networks that use frequency bins together with the polarization degree of freedom, either to encode information or as a resource to distribute entanglement, active stabilization of the PDPM is an important next step.

\bibliography{pdpm_ol_refs}

\begin{thebibliography}{10}
\newcommand{\enquote}[1]{``#1''}

\bibitem{Barreiro2008}
J.~T. Barreiro, T.-C. Wei, and P.~G. Kwiat, \enquote{{Beating the channel
  capacity limit for linear photonic superdense coding},}
  {\protect\JournalTitle{Nature Physics}} \textbf{4}, 282--286 (2008).

\bibitem{graham2015}
T.~M. Graham, H.~J. Bernstein, T.-C. Wei, M.~Junge, and P.~G. Kwiat,
  \enquote{Superdense teleportation using hyperentangled photons,}
  {\protect\JournalTitle{Nature communications}} \textbf{6}, 7185 (2015).

\bibitem{Walborn2003}
S.~P. Walborn, S.~P{\'{a}}dua, and C.~H. Monken,
  \enquote{{Hyperentanglement-assisted Bell-state analysis},}
  {\protect\JournalTitle{Physical Review A}} \textbf{68}, 042313 (2003).

\bibitem{Schuck2006}
C.~Schuck, G.~Huber, C.~Kurtsiefer, and H.~Weinfurter, \enquote{{Complete
  Deterministic Linear Optics Bell State Analysis},}
  {\protect\JournalTitle{Physical Review Letters}} \textbf{96}, 190501 (2006).

\bibitem{lukens2017}
J.~M. Lukens and P.~Lougovski, \enquote{Frequency-encoded photonic qubits for
  scalable quantum information processing,} {\protect\JournalTitle{Optica}}
  \textbf{4}, 8--16 (2017).

\bibitem{Kues2017}
M.~Kues, C.~Reimer, P.~Roztocki, L.~R. Cort{\'{e}}s, S.~Sciara, B.~Wetzel,
  Y.~Zhang, A.~Cino, S.~T. Chu, B.~E. Little, D.~J. Moss, L.~Caspani,
  J.~Aza{\~{n}}a, and R.~Morandotti, \enquote{{On-chip generation of
  high-dimensional entangled quantum states and their coherent control},}
  {\protect\JournalTitle{Nature}} \textbf{546}, 622--626 (2017).

\bibitem{imany2018}
P.~Imany, J.~A. Jaramillo-Villegas, O.~D. Odele, K.~Han, D.~E. Leaird, J.~M.
  Lukens, P.~Lougovski, M.~Qi, and A.~M. Weiner, \enquote{50-ghz-spaced comb of
  high-dimensional frequency-bin entangled photons from an on-chip silicon
  nitride microresonator,} {\protect\JournalTitle{Optics Express}} \textbf{26},
  1825--1840 (2018).

\bibitem{Imany2018pra}
P.~Imany, O.~D. Odele, J.~A. Jaramillo-Villegas, D.~E. Leaird, and A.~M.
  Weiner, \enquote{{Characterization of coherent quantum frequency combs using
  electro-optic phase modulation},} {\protect\JournalTitle{Physical Review A}}
  \textbf{97}, 1--5 (2018).

\bibitem{wooten2000}
E.~L. Wooten, K.~M. Kissa, A.~Yi-Yan, E.~J. Murphy, D.~A. Lafaw, P.~F.
  Hallemeier, D.~Maack, D.~V. Attanasio, D.~J. Fritz, G.~J. McBrien
  \emph{et~al.}, \enquote{A review of lithium niobate modulators for
  fiber-optic communications systems,} {\protect\JournalTitle{IEEE Journal of
  selected topics in Quantum Electronics}} \textbf{6}, 69--82 (2000).

\bibitem{wang2018}
C.~Wang, M.~Zhang, X.~Chen, M.~Bertrand, A.~Shams-Ansari, S.~Chandrasekhar,
  P.~Winzer, and M.~Lon{\v{c}}ar, \enquote{Integrated lithium niobate
  electro-optic modulators operating at cmos-compatible voltages,}
  {\protect\JournalTitle{Nature}} p.~1 (2018).

\bibitem{fukuda2008}
H.~Fukuda, K.~Yamada, T.~Tsuchizawa, T.~Watanabe, H.~Shinojima, and S.-i.
  Itabashi, \enquote{Silicon photonic circuit with polarization diversity,}
  {\protect\JournalTitle{Optics express}} \textbf{16}, 4872--4880 (2008).

\bibitem{haffner2018}
C.~Haffner, D.~Chelladurai, Y.~Fedoryshyn, A.~Josten, B.~Baeuerle, W.~Heni,
  T.~Watanabe, T.~Cui, B.~Cheng, S.~Saha \emph{et~al.}, \enquote{Low-loss
  plasmon-assisted electro-optic modulator,} {\protect\JournalTitle{Nature}}
  \textbf{556}, 483 (2018).

\bibitem{Brodsky2011}
M.~Brodsky, E.~C. George, C.~Antonelli, and M.~Shtaif, \enquote{{Loss of
  polarization entanglement in a fiber-optic system with polarization mode
  dispersion in one optical path},} {\protect\JournalTitle{Optics Letters}}
  \textbf{36}, 43--45 (2011).

\bibitem{Jones2018}
D.~E. Jones, B.~T. Kirby, and M.~Brodsky, \enquote{{Tuning quantum channels to
  maximize polarization entanglement for telecom photon pairs},}
  {\protect\JournalTitle{npj Quantum Information}} \textbf{4}, 58 (2018).

\bibitem{weiner2011}
A.~Weiner, \emph{Ultrafast optics}, vol.~72 (John Wiley \& Sons, 2011).

\end{thebibliography}

\bibliographyfullrefs{pdpm_ol_refs}

\end{document}